\documentclass[11pt]{article}
\usepackage[dvips]{graphicx}
\usepackage{mathrsfs}
\usepackage{eucal}
\usepackage{amsmath,amsthm,amssymb}
\usepackage{wrapfig}
\usepackage[dvips]{color}
\usepackage{cite}
\usepackage{framed}
\usepackage{tabularx}
\usepackage[sectionbib]{chapterbib}
\usepackage{threeparttable}
\usepackage{float}
\definecolor{shadecolor}{gray}{0.80}
\DeclareMathVersion{chem}
\SetSymbolFont{letters}{chem}{OT1}{cmr}{m}{n}
\definecolor{shadecolor}{rgb}{0.9, 0.9, 0.90}
\newcommand{\HG}{\hspace{1.3mm}\Hat{\textit{\textsf{\hspace{-1.3mm}G}}}\hspace{0.5mm}}

\begin{document}
\renewcommand{\figurename}{\small{Fig.}~}
\renewcommand{\thefootnote}{$\dagger$\arabic{footnote}}

\begin{flushright}
\textit{Theory of Excluded Volume Effect}
\end{flushright}

\vspace{1mm}
\begin{center}
\setlength{\baselineskip}{25pt}{\LARGE\textbf{Alternative Approach to the Excluded Volume Problem}}
\end{center}
\vspace{-7mm}
\begin{center}
\setlength{\baselineskip}{25pt}{\large\textbf{The Critical Behavior of the Exponent $\nu$}}
\end{center}

\vspace*{4mm}
\begin{center}
\large{Kazumi Suematsu\footnote{The author takes full responsibility for this article.}, Haruo Ogura\footnote{Kitasato University}, Seiiti Inayama\footnote{Keio University} and Toshihiko Okamoto\footnote{Tokyo University}} \vspace*{2mm}\\
\normalsize{\setlength{\baselineskip}{12pt}
Institute of Mathematical Science\\
Ohkadai 2-31-9, Yokkaichi, Mie 512-1216, JAPAN\\
E-Mail: ksuematsu@icloud.com,  Tel/Fax: +81 (0) 59 326 8052}\\[8mm]
\end{center}

\vspace{3mm}
%%%%%%%%%%%%%%%%%%
\hrule
\vspace{4mm}
\noindent\textbf{\large Abstract}\\[2mm]
We present the alternative derivation of the excluded volume equation. The resulting equation is mathematically identical to the one proposed in the preceding paper. As a result, the theory reproduces well the observed points by SANS (small angle neutron scattering) experiments. The equation is applied to the coil-globule transition of branched molecules. It is found that in the entire region of poor solvent regimes ($T<\Theta$), the exponent $\kappa=d\log\alpha/d\log N\, (N\rightarrow\infty)$ takes the value $\frac{1}{12}$, showing that contrary to the case of linear molecules ($\kappa=-\frac{1}{6}$), the expansion factor increases indefinitely as $N$ increases. The theory is then applied to concentrated systems in good solvents. It is found that for the entire region of $0<\bar{\phi}\le 1$, the gradients $\kappa$ seem to converge on a common value lying somewhere from $\kappa=\frac{1}{12}$ to $0.1$. Since $\nu_{dilute}=\tfrac{1}{2}$, $\nu_{melt}=\tfrac{1}{3}$,  and $0.33\cdots\le\nu_{conc}\,(=\nu_{0}+\kappa) <0.35$ for $0<\bar{\phi}\le 1$, the simulation results suggest that the exponents $\kappa$ and $\nu$ change abruptly from phases to phases; there are no intermediate values between them, for instance between $\nu_{dilute}$ and $\nu_{melt}$.

\vspace{3mm}
\noindent\textbf{Key Words}:
\normalsize{Excluded Volume Effects/ Alternative Derivation/ Critical Nature of the Exponent/ Coil-Globule Transition}\\[0mm]

\hrule
\vspace{3mm}
\setlength{\baselineskip}{13pt}
%%%%%%%%%%%%%%%%%% Introduction
\section{Introduction}
The classic theory\cite{Flory} of the excluded volume effects has been constructed on the sound physical basis\cite{Kazumi}, whereas the theory is strictly restricted to the description of the dilution limit. Recently, the theory was generalized so as to describe the entire range of the polymer concentration. In  applying the theory to real problems, on the other hand, we must introduce some approximations. A major one is the Gaussian approximation of segment distribution around the center of gravity. It is known that, even though the difference from the true distribution\cite{Mazur} is by no means serious, the Gaussian approximation is not exact both for the unperturbed chain\cite{Ishihara, Debye} and for the expanded coil\cite{Mazur}. Meanwhile, it was found that conclusions deduced from this approximation reproduce well experimental points and make predictions of new phenomena\cite{Kazumi}. Encouraged by this success, we have applied, in the preceding work, the same Gaussian distribution to branched polymers and estimated the exponent, $\nu_{melt}=\frac{1}{3}$ for the radius of gyration. While the result seems reasonable, the underlying theoretical construction was not mathematically consistent. In this paper, we will concentrate on mathematical consistency, showing that the same formula can be derived from a new thermodynamic point of view. Making use of the revised theory we approach the riddle of the critical nature of the exponent, $\nu$.
%%%%%%%%%%%%%%%%%% Section 1
\section{Theoretical}
The fundamental equation of the Gibbs potential of mixing pure solvent and pure polymer melt:
%%%%%%%%%%%%%%%%%% Eq. 1
\begin{equation}
\Delta G_{mixing}=\,\frac{kT}{V_{1}}\int\left\{-\left(1-\chi\right)v_{2}+\left(1/2-\chi\right)v_{2}^{2}+\frac{1}{6}v_{2}^{3}+\cdots\right\}\delta V\label{AlD-1}
\end{equation}
where $V$ denotes the system volume and $V_{1}$ the volume of a solvent molecule, and $v_{2}$ represents the volume fraction of polymer segments in the local area $\delta V$. Let $V_{2}$ denote the volume of a segment. Assuming the Gaussian distribution of segments around the center of gravity, $v_{2}$ can be expressed in the form:
%%%%%%%%%%%%%%%%%% Eq. 2
\begin{align}
\Hat{v}_{2}=&V_{2}\Hat{C}=V_{2}N\left(\frac{\beta}{\pi\alpha^{2}}\right)^{3/2}\sum_{\{a, b, c\}}\exp\left\{-\frac{\beta}{\alpha^{2}}\left[(x-a)^{2}+(y-b)^{2}+(z-c)^{2}\right]\right\}\notag\\
\equiv&V_{2}N\left(\frac{\beta}{\pi\alpha^{2}}\right)^{3/2}\HG(x, y, z)\label{AlD-2}
\end{align}
with $\beta=3/2\langle s_{N}^{2}\rangle_{0}$ having the usual meaning (the subscript 0 denotes the unperturbed dimensions), and $\{a, b, c\}$ denoting the location of individual polymer molecules, so that $\Hat{C}$ represents the number concentration of segments at the coordinate $(x, y, z)$ and the symbol \,$\Hat{}$\, means the Gaussian approximation specified by Eq. (\ref{AlD-2}). 

Let us return to the original thermodynamic equation. Let $n_{i}$ be the number of molecules of species $i$. The basic thermodynamic equation is
%%%%%%%%%%%%%%%%%% Eq. 3
\begin{equation}
d G=V dP- S dT+\sum_{i}\left(\frac{\partial G}{\partial n_{i}}\right)_{T,P}dn_{i}\label{AlD-3}
\end{equation}
Multiplying Eq. (\ref{AlD-3}) by $V/V$ (or by the ratio of local volume, $\delta V/\delta V$), we have
%%%%%%%%%%%%%%%%%% Eq. 4
\begin{equation}
d G=V dP- S dT+\sum_{i}\left(\frac{\partial G}{\partial c_{i}}\right)_{T,P}dc_{i}\label{AlD-4}
\end{equation}
where $c_{i}$ represents the number concentration of molecular species $i$. Thus we can introduce a new definition of the chemical potential as a measure of the rate of the change of Gibbs potential as against the change of solute concentration\cite{deGennes} under constant $T$ and $P$:
%%%%%%%%%%%%%%%%%% Eq. 5
\begin{equation}
\mu_{c_{i}}=\left(\frac{\partial G}{\partial c_{i}}\right)_{T,P}\label{AlD-5}
\end{equation}
Then let us apply Eq. (\ref{AlD-5}) to the free energy of mixing, $\Delta G_{mixing}$, to get
%%%%%%%%%%%%%%%%%% Eq. 6
\begin{equation}
\Delta \mu_{mixing}=\left(\frac{\partial\Delta G_{mixing}}{\partial \Hat{C}}\right)_{T,P}=\left\{\left(\frac{\partial\Delta G_{mixing}}{\partial \Hat{v}_{2}}\right)\left(\frac{\partial \Hat{v}_{2}}{\partial \Hat{C}}\right)\right\}_{T,P}\label{AlD-6}
\end{equation}
Since
%%%%%%%%%%%%%%%%%% Eq. 7
\begin{equation}
\left(\frac{\partial \Hat{v}_{2}}{\partial \Hat{C}}\right)=V_{2}\label{AlD-7}
\end{equation}
we have
%%%%%%%%%%%%%%%%%% Eq. 8
\begin{equation}
\Delta \mu_{mixing}=\,kT\frac{V_{2}}{V_{1}}\int\left\{-\left(1-\chi\right)+\left(1-2\chi\right)\Hat{v}_{2}+\frac{1}{2}\Hat{v}_{2}^{2}+\cdots\right\}\delta V\label{AlD-8}
\end{equation}
By taking the difference from the pure solvent ($\Hat{v}_{2}=0$), Eq. (\ref{AlD-8}) may be recast in the form:
%%%%%%%%%%%%%%%%%% Eq. 9
\begin{equation}
\Delta\mu_{\Hat{v}_{2}}=\Delta\mu_{0}+\,kT\frac{V_{2}}{V_{1}}\int\left\{\left(1-2\chi\right)\Hat{v}_{2}+\frac{1}{2}\Hat{v}_{2}^{2}+\cdots\right\}\delta V\label{AlD-9}
\end{equation}
where
%%%%%%%%%%%%%%%%%%
\begin{equation}
\Delta\mu_{0}=\,kT\frac{V_{2}}{V_{1}}\int\bigg\{-\left(1-\chi\right)\bigg\}\delta V\notag
\end{equation}
is independent of the segment volume fraction $\Hat{v}_{2}$.
A central theme is to evaluate the potential difference between the inside of a polymer molecule and the outside. Let $\Hat{v}_{2, hill}$ represent the volume fraction of segments in the inside and $\Hat{v}_{2, valley}$ that in the outside. The difference in the chemical potential is directly related to the movement of segments from a more concentrated ($hill$) region to a more dilute ($valley$) region. It has the form:
%%%%%%%%%%%%%%%%%% Eq. 10
\begin{align}
\Delta \mu_{osmotic}=&\,\Delta \mu_{\Hat{v}_{2}, hill}-\Delta \mu_{\Hat{v}_{2}, valley}\notag\\
=&\,kT\frac{V_{2}}{V_{1}}\int\left\{\left(1-2\chi\right)\Hat{\mathscr{J}}^{1}+\frac{1}{2}\Hat{\mathscr{J}}^{2}+\cdots\right\}\delta V\label{AlD-10}
\end{align}
where $\Hat{\mathscr{J}}^{k}=\Hat{v}_{2, hill}^{k}-\Hat{v}_{2, valley}^{k}$. As it turns out, the excluded volume problem belongs to a science that deals with the osmosis of segments. In contrast to the case of the homogeneous solution of small solutes, a polymer solution cannot attain the equilibrium state through $\Delta \mu_{osmotic}$ alone. This is because monomers are joined by chemical bonds. There is retraction force due to the elastic potential of a polymer coil, requiring the other equilibrium condition that can be realized at a point where the force of the osmosis counterbalances the force of the elasticity. The Gibbs potential in the system is thus $\Delta G_{sys}=\Delta G_{osmotic}+\Delta G_{elastic}$. Our final end is to find the solution of the equality:
%%%%%%%%%%%%%%%%%% Eq. 11
\begin{equation}
\left(\frac{\partial\Delta G_{sys}}{\partial \alpha}\right)_{T,P}=\left(\frac{\partial\Delta G_{osmotic}}{\partial \alpha}\right)_{T,P}+\left(\frac{\partial\Delta G_{elastic}}{\partial \alpha}\right)_{T,P}=0\label{AlD-11}
\end{equation}
Since, $\left(\partial\Delta G_{osmotic}/\partial \alpha\right)_{T,P}=\Delta \mu_{osmotic}\,(\partial\Hat{C}/\partial\alpha)$, the first term on the rhs of Eq. (\ref{AlD-11}) can be calculated as:
%%%%%%%%%%%%%%%%%% Eq. 12
\begin{multline}
\left(\frac{\partial\Delta G_{osmotic}}{\partial \alpha}\right)_{T,P}=kT\frac{V_{2}}{V_{1}}\left\{\iiint\left(\left(1-2\chi\right)\Hat{v}_{2, hill}+\frac{1}{2}\Hat{v}_{2, hill}^{2}+\cdots\right)\left(\frac{\partial\Hat{C}}{\partial \alpha}\right)_{\hspace{-0.3mm}hill} dx dy dz\right.\\
\left.-\iiint\left(\left(1-2\chi\right)\Hat{v}_{2, valley}+\frac{1}{2}\Hat{v}_{2, valley}^{2}+\cdots\right)\left(\frac{\partial\Hat{C}}{\partial \alpha}\right)_{\hspace{-0.3mm}valley} dx dy dz\right\}\label{AlD-12}
\end{multline}
In Eq. (\ref{AlD-12}), the subscripts, $hill$ and $valley$, are only formal description. The differentiation of $\Hat{C}$ is carried out making use of Eq. (\ref{AlD-2}), equally for both $hill$ and $valley$. The mathematical difference between $hill$ and $valley$ arises only through the integral operation. Hence
%%%%%%%%%%%%%%%%%% Eq. 13
\begin{equation}
\left(\frac{\partial\Hat{C}}{\partial \alpha}\right)=-\frac{3}{\alpha}\frac{\Hat{v}_{2}}{V_{2}}+\frac{2\beta N}{\alpha^{3}}\left(\frac{\beta}{\pi\alpha^{2}}\right)^{3/2}\sum_{\{a, b, c\}}s^{2}(x, y, z, a, b, c)\exp\left\{-\frac{\beta}{\alpha^{2}}s^{2}(x, y, z, a, b, c)\right\}\label{AlD-13}
\end{equation}
where $s^{2}(x, y, z, a, b, c)=(x-a)^{2}+(y-b)^{2}+(z-c)^{2}$.
The expression for the elastic potential is already given in the previous works \cite{Flory, Treloar, Kazumi} and has the form:
%%%%%%%%%%%%%%%%%% Eq. 14
\begin{equation}
\left(\frac{\partial\Delta G_{elastic}}{\partial \alpha}\right)_{T,P}=-T\left(\frac{\partial \Delta S}{\partial \alpha}\right)_{T, P}=3kT\left(\alpha-1/\alpha\right)\label{AlD-14}
\end{equation}
Substituting Eqs. (\ref{AlD-12}) and (\ref{AlD-14}) into Eq. (\ref{AlD-11}), we have finally
%%%%%%%%%%%%%%%%%% Eq. 15
\begin{multline}
\alpha-1/\alpha =-\frac{V_{2}}{3V_{1}}\left\{\iiint\left(\left(1-2\chi\right)\Hat{v}_{2, hill}+\frac{1}{2}\Hat{v}_{2, hill}^{2}+\cdots\right)\left(\frac{\partial\Hat{C}}{\partial \alpha}\right)_{\hspace{-0.3mm}hill} dx dy dz\right.\\
\left.-\iiint\left(\left(1-2\chi\right)\Hat{v}_{2, valley}+\frac{1}{2}\Hat{v}_{2, valley}^{2}+\cdots\right)\left(\frac{\partial\Hat{C}}{\partial \alpha}\right)_{\hspace{-0.3mm}valley} dx dy dz\right\}\label{AlD-15}
\end{multline}
which may be recast in the form:
%%%%%%%%%%%%%%%%%% Eq. 16
\begin{equation}
\alpha-1/\alpha =-\frac{V_{2}}{3V_{1}}\iiint\left\{\left(1-2\chi\right)\Hat{\mathscr{J}}^{1}+\frac{1}{2}\Hat{\mathscr{J}}^{2}+\cdots\right\}\left(\frac{\partial\Hat{C}}{\partial \alpha}\right) dx dy dz\label{AlD-16}
\end{equation}
where $\Hat{\mathscr{J}}^{k}=\Hat{v}_{2, hill}^{k}-\Hat{v}_{2, valley}^{k}$ as defined above. One feature of the present derivation is that the first term in the potential function, Eq. (\ref{AlD-1}), which is superfluous in the theory of the excluded volume effects, can be removed in a natural fashion from the final expression. In the present problem, $dV = dxdydz$ and $d\alpha$ are independent operators, so that the change in order of the integral and the differentiation does not affect the result. So, as it turns out, Eq. (\ref{AlD-16}) is mathematically equivalent to the corresponding equality in the preceding paper\cite{Kazumi}:
%%%%%%%%%%%%%%%%%% Eq. 17
\begin{equation}
\alpha-1/\alpha =-\frac{1}{3V_{1}}\frac{\partial}{\partial\alpha}\iiint\left\{\left(1/2-\chi\right)\Hat{\mathscr{J}}^{2}+\frac{1}{6}\Hat{\mathscr{J}}^{3}+\cdots\right\}dxdydz\label{AlD-17}
\end{equation}

Unfortunately, except for the dilution limit, Eqs. (\ref{AlD-15})-(\ref{AlD-17}) don't appear to yield neat solutions, since the integration interval corresponding to ``$hill$'' and ``valley'' cannot be specified. However we can extract the essential features of these equations by making use of the lattice model\cite{Kazumi} with the help of Eqs. (\ref{AlD-2}) and (\ref{AlD-13}), as functions of $\alpha$, $N$ and the polymer concentration. Polymers are put on the sites of the simple cubic lattice, whose dimensions are $p\times p\times p$, with the Gaussian distribution. The mean segment fraction ($\bar{\phi}$) can be calculated using $p$ by the equality: $\bar{\phi}=V_{2}N/p^{3}$.

\begin{figure}[H]
\begin{center}
\includegraphics[width=9cm]{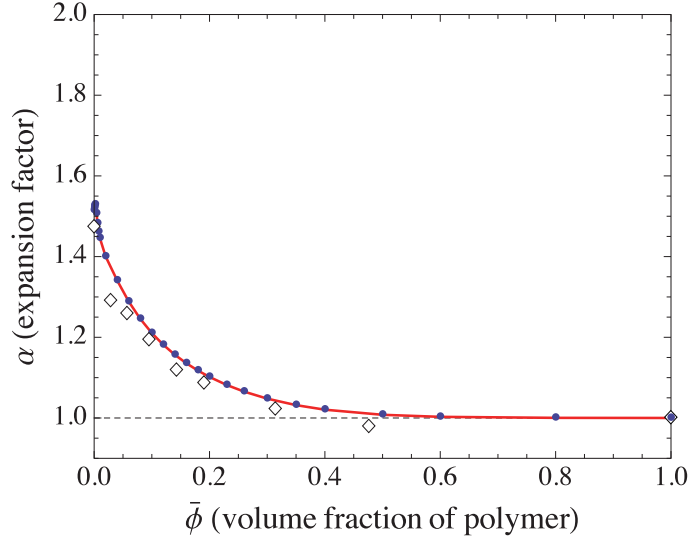}
\caption{Expansion factor vs. $\bar{\phi}$ plot for PSt$-$CS$_{2}$. Filled circles ($\bullet$): theoretical points by Eq. (\ref{AlD-18}) for $\chi=0.4$; diamonds ($\lozenge$): observed points by Daoud and coworkers (the observed value of 82 \text{\AA} at $\bar{\phi}=1$ has been replaced by the revised one 93 \text{\AA})\cite{Daoud}.}\label{PSt-CS2}
\end{center}
\end{figure}

Before applying Eq. (\ref{AlD-15}) to real problems, it is necessary to confirm the sound basis of the theory through comparison with experimental observations. For this purpose, we use the reduced form of Eq. (\ref{AlD-15}):
%%%%%%%%%%%%%%%%%% Eq. 18
\begin{multline}
\alpha-1/\alpha =-\frac{V_{2}}{3V_{1}}\left\{\iiint_{-p/4}^{p/4}\left(1-2\chi\right)\Hat{v}_{2, hill}\left(\frac{\partial\Hat{C}}{\partial \alpha}\right)_{\hspace{-0.3mm}hill} dx dy dz\right.\\
-\left.\iiint_{p/4}^{3p/4}\left(1-2\chi\right)\Hat{v}_{2, valley}\left(\frac{\partial\Hat{C}}{\partial \alpha}\right)_{\hspace{-0.3mm}valley} dx dy dz\right\}\label{AlD-18}
\end{multline}
The numerical analysis of Eq. (\ref{AlD-18}) is carried out taking the integration interval from $-p/4$ to $p/4$ for each axis for the $hill$ area, and from $p/4$ to $3p/4$ for the $valley$ area. A simulation result that models the polystyrene(PSt)$-$CS$_{2}$ system is displayed in Fig. \ref{PSt-CS2} to be compared with the SANS experiments by Daoud and coworkers\cite{Daoud}. It is seen that the resulting $\alpha$ vs.\,$\bar{\phi}$ curve ($\bullet$), which is exactly the same as calculated previously using Eq. (\ref{AlD-17})\cite{Kazumi}, reproduces well the observed points ($\lozenge$).
%%%%%%%%%%%%%%%%%% Fig. 1

Having confirmed the sound basis of the theory, we proceed to the application of the theory to branched polymers. Let us begin by the examination of the behavior in the poor solvent region $(1/2-\chi<0)$, or equivalently $T<\Theta$ region. 

\subsection{Branched Polymers in $T<\Theta$ Region}
It is well-established that  linear molecules obey the coil-globule transition in the poor solvent regime\cite{Lifshitz, Sanchez, Sun}. Hence it may be of interest to investigate the corresponding behavior of  the branched polymers under the same conditions. Since the system often encounters the phase separation in poor solvents, it will be reasonable to inspect the dilution limit, $\bar{\phi}\rightarrow 0$.

Let the system be comprised of a single branched molecule and a large amount of the solvent. Then we have $\Hat{v}_{2, valley}=0$ and Eq. (\ref{AlD-2}) reduces to
$$\Hat{v}_{2, hill}=V_{2}N\left(\frac{\beta}{\pi\alpha^{2}}\right)^{\frac{3}{2}}\exp\left\{-\frac{\beta}{\alpha^{2}}\left(x^{2}+y^{2}+z^{2}\right)\right\}.$$
In this case, the integration interval in Eq. (\ref{AlD-15}) must be taken from $-\infty$ to $\infty$, then Eq. (\ref{AlD-15}) yields the known equality:
%%%%%%%%%%%%%%%%%% Eq. 19
\begin{equation}
\alpha^{5}-\alpha^{3}=N^{2}\frac{V_{2}^{\,2}}{V_{1}}\left(\frac{\beta}{\pi}\right)^{\frac{3}{2}}\left\{\frac{1}{2^{\frac{3}{2}}}\left(1/2-\chi\right)+\frac{V_{2}N}{3^{\frac{3}{2}+1}\alpha^{3}}\left(\frac{\beta}{\pi}\right)^{\frac{3}{2}}+\cdots\right\}\label{AlD-19}
\end{equation}
Only difference from linear polymers is the radius of gyration\cite{Zim, Dobson, Kajiwara, deGennes, Kazumi}:
%%%%%%%%%%%%%%%%%% Eq. 20
\begin{equation}
\begin{aligned}
\langle s_{N}^{2}\rangle_{0}&=\frac{1}{2N^{2}}\,\frac{\displaystyle N! \{(f-2)N+2\}!}{\displaystyle\{(f-1)N\}!}\sum_{k=1}^{N-1}\binom{(f-1)k}{k-1}\binom{(f-1)(N-k)}{N-k-1}l^{2}\\
&\simeq \left(\frac{(f-1)\pi}{2^{3}(f-2)}\right)^{1/2} N^{\frac{1}{2}}l^{2}\hspace{5mm}(\text{as}\,\, N\rightarrow \infty)
\end{aligned}\label{AlD-20}
\end{equation}

Following the preceding paper, we neglect the terms higher than the third to obtain
%%%%%%%%%%%%%%%%%% Eq. 19'
\begin{equation}
\alpha^{5}-\alpha^{3}=N^{2}\frac{V_{2}^{\,2}}{V_{1}}\left(\frac{\beta}{\pi}\right)^{\frac{3}{2}}\left\{\frac{1}{2^{\frac{3}{2}}}\left(1/2-\chi\right)+\frac{V_{2}N}{3^{\frac{3}{2}+1}\alpha^{3}}\left(\frac{\beta}{\pi}\right)^{\frac{3}{2}}\right\}\tag{\ref{AlD-19}'}
\end{equation}
Let us solve Eq. (\ref{AlD-19}') using parameters shown in Table \ref{AlD-Table1}. An important point is that a branched polymer is immersed in the solvent with large molecular volume $\left(V_{1}=10^{4}\times V_{2}\right)$. From a theoretical standpoint, this is a convenient assumption, since as $N$ increases, the volume ratio of a polymer molecule to the solvent is reversed, and in the event, the system enters the regime of $V_{2}N\gg V_{1}$, in which the system becomes equivalent to the ordinary polymer solution, i.e., a large polymer molecule immersed in the small solvent molecules.

%%%%%%%%%%%%%%%%%% Table 1
\begin{center}
  \begin{threeparttable}[h]
    \caption{Parameters of a hypothetical branched polymer solution ($d=3$)}\label{AlD-Table1}
  \begin{tabular}{l l c r}
\hline\\[-1.5mm]
& \hspace{10mm}parameters & notations & values \,\,\,\,\\[2mm]
\hline\\[-1.5mm]
solvent & volume of a solvent & $V_{1}$ & \hspace{5mm}$10^{4}\times V_{2}$ \text{\AA}$^{3}$\\[1.5mm]
branched polymer & volume of a segment & $V_{2}$ & \hspace{5mm}387 \text{\AA}$^{3}$\\[1.5mm]
& mean bond length & $\bar{l}$ & \hspace{5mm}10 \text{\AA}\,\,\,\\[1.5mm]
& enthalpy parameter & $\chi$ & $>1/2$\hspace{3.6mm} \,\,\\[2mm]
\hline\\[-6mm]
   \end{tabular}
    \vspace*{2mm}
  \end{threeparttable}
  \vspace*{4mm}
\end{center}

\begin{figure}[h]
\begin{center}
\begin{minipage}[t]{0.46\textwidth}
\begin{center}
\includegraphics[width=7.6cm]{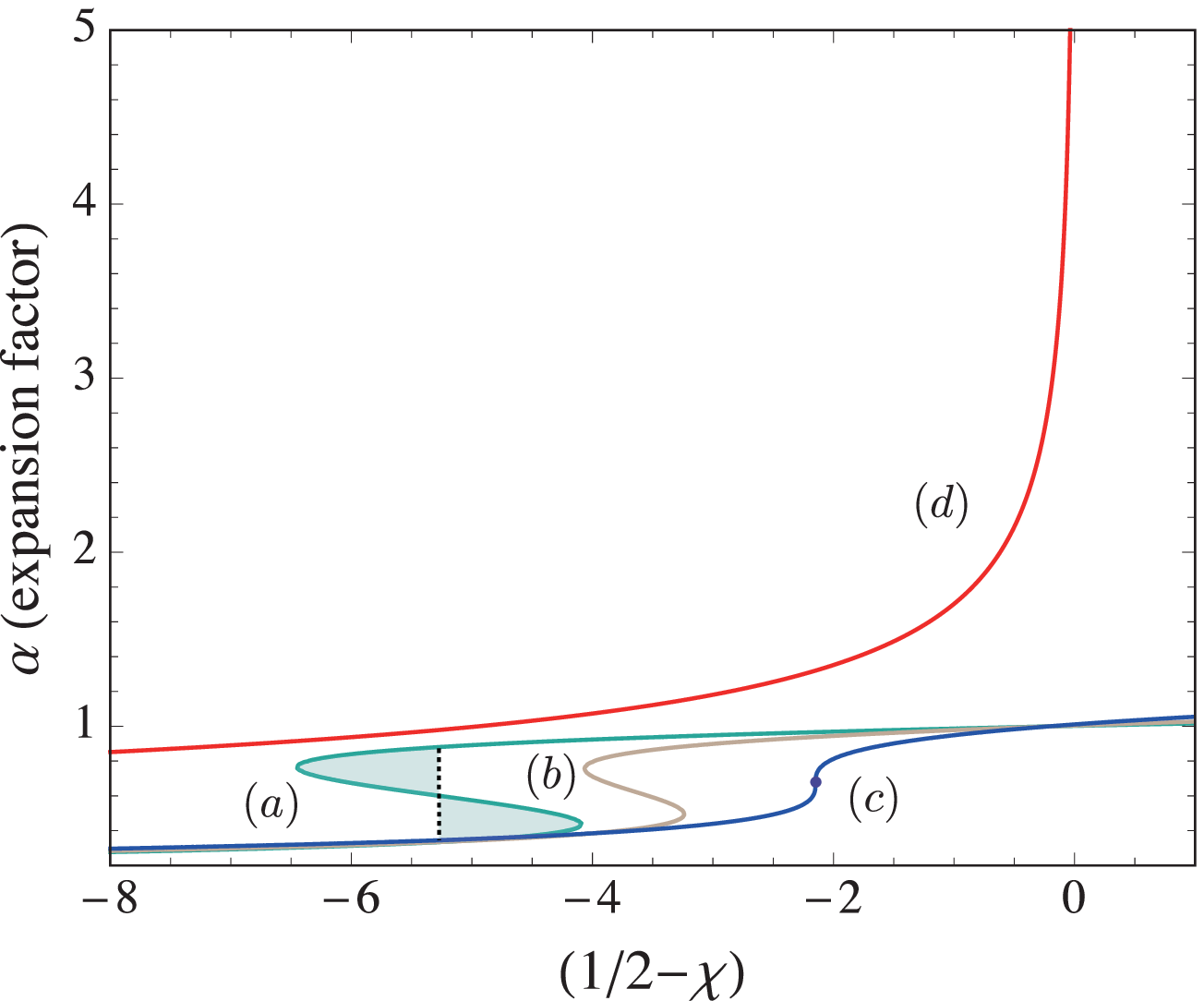}
\end{center}
\vspace{-2mm}
\caption{Expansion factor vs. $(1/2-\chi)$ plot for the hypothetical branched polymer solution ($d=3$). A single branched molecule immersed in large solvent molecules having $V_{1}=10^{4}\, V_{2}$. Solid lines are theoretical ones by Eq. (\ref{AlD-19}') for (a) $N=1000$, (b) $N=1500$, (c) $N=2843$ and (d) $N=10^{9}$.}\label{SigmoidBP}
\end{minipage}
\hspace{10mm}
%%%%%%%%%%%%%%%%%% Fig-3
\begin{minipage}[t]{0.46\textwidth}
\begin{center}
\includegraphics[width=7.8cm]{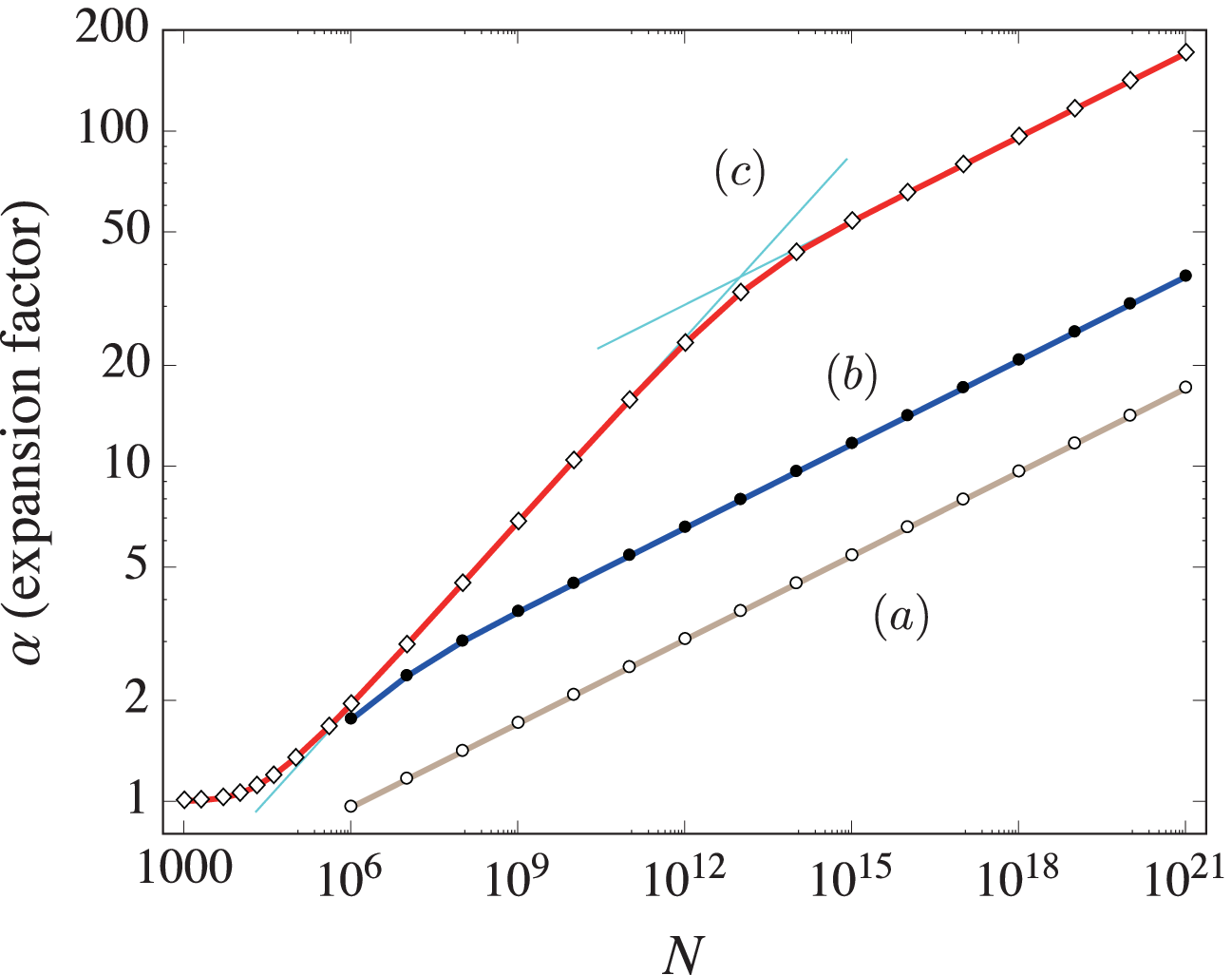}
\end{center}
\vspace{-2mm}
\caption{Expansion factor vs. $N$ plot for the hypothetical branched polymer system. Calculated by Eq. (\ref{AlD-19}') for $(a)$ $(1/2-\chi)=-1$, $(b)$ $(1/2-\chi)=-0.1$ and $(c)$ $(1/2-\chi)=-0.001$. For all the cases, $\kappa=\frac{1}{12}$ as $N\rightarrow\text{large}$.}\label{C-GExponent}
\end{minipage}
\end{center}
\vspace*{-4mm}
\end{figure}

Under the above conditions, we have performed the simulation of Eq. (\ref{AlD-19}') for $1/2-\chi<0$, with varying degree, $N$, of polymerization. Typical examples are illustrated in Fig. \ref{SigmoidBP}. Let us define the exponent $\kappa$ by the relation $\alpha\propto N^{\kappa}$ ($N\rightarrow\infty$). It is seen from Fig. \ref{SigmoidBP} that contrary to the case of linear polymers ($\kappa=-\frac{1}{6}$), $\alpha$ is an increasing function of $N$, namely $\kappa>0$. The sigmoid curves observed for smaller molecules ($N\le 2843$) disappear with increasing $N$, eventually leading to a monotonic function represented by the curve $(d)$. 

%%%%%%%%%%%%%%%%%% Fig-2
Whether a polymer is linear or branched, the problem of the coil-globule transition reduces to the problem of seeking the roots of the polynomial equation of the form:
%%%%%%%%%%%%%%%%%% Eq. 21
\begin{equation}
H(\alpha)=\alpha^{5}-\alpha^{3}-\Big\{c_{1}\left(\tfrac{1}{2}-\chi\right)+c_{2}/\alpha^{3}\Big\}=0\label{AlD-21}
\end{equation}
Let us use the abbreviation: $\left\{c_{1}\left(\tfrac{1}{2}-\chi\right)+c_{2}/\alpha^{3}\right\}\equiv \{\cdots\}$. In order for Eq. (\ref{AlD-21}) to have three positive roots in the interval $0<\alpha<1$, namely to have the sigmoid curves shown in Fig. \ref{SigmoidBP}, the term $\{\cdots\}$ must take a small minus value, because $\alpha^{5}-\alpha^{3}<0$ when $0<\alpha<1$.  The simulation condition, ``a small branched polymer immersed in the solvent with the large molecular volume'', is simply a mathematical requirement to realize such circumstances. 

In Fig. \ref{C-GExponent}, the $N$ dependence of $\alpha$ is illustrated. It is readily noticed that the curve $(c)$ ($1/2-\chi=-0.001$) has an inflection point at $N\approx 10^{13}$. The first linear region ($N<10^{13}$) on this curve has the slope of $\kappa =\frac{3}{16}$, so it corresponds to the exponent of the $\Theta$ region, while the second region of $N>10^{13}$ has the slope of $\kappa =\frac{1}{12}$ corresponding to the globular form. Since $\langle s_{N}^{2}\rangle^{1/2}=\alpha\langle s_{N}^{2}\rangle^{1/2}_{0}\propto N^{\nu}$, it follows that $\nu=\kappa+\nu_{0}$, from which using $\nu_{0}=\frac{1}{4}$ for unperturbed branched polymers, we have $\nu_{globule}=\frac{1}{3}$. This is equal to the critical packing density. It is seen that identically to the case of linear polymers\cite{deGennes, Lifshitz}, the globule of a branched polymer having a large molecular weight is in the liquid or the solid state.\\[-3mm]

The exponents $\kappa$ and $\nu$ may be estimated through intuitive discussion. Since $\alpha\propto N^{\kappa}$ ($N\rightarrow\infty$) by definition, we may recast Eq. (\ref{AlD-19}') for the $T<\Theta$ region in the scaling form:
%%%%%%%%%%%%%%%%%% Eq. 22
\begin{equation}
\alpha^{5}-\alpha^{3}\sim -N^{2-3\nu_{0}}+N^{3-6\nu_{0}-3\kappa}\label{AlD-22}
\end{equation}
Firstly, we must have $2-3\nu_{0}\le 3-6\nu_{0}-3\kappa$, because otherwise Eq. (\ref{AlD-22}) gives $\alpha<1$, which contradicts the critical packing density criterion $\nu\ge\frac{1}{3}$. Secondly we must have $2-3\nu_{0}\ge 3-6\nu_{0}-3\kappa$, because otherwise the convergency of the series expansion in Eq. (\ref{AlD-1}) cannot be fulfilled. In order for the two opposite requirements to be reconciled, we must have $2-3\nu_{0} =3-6\nu_{0}-3\kappa$, giving
%%%%%%%%%%%%%%%%%% Eq. 23
\begin{equation}
\kappa=\frac{1-3\hspace{0.3mm}\nu_{0}}{3}\label{AlD-23}
\end{equation}
Substituting $\nu_{0}=\frac{1}{4}$ into Eq. (\ref{AlD-23}), we recover the foregoing exponents $\kappa=\frac{1}{12}$ and $\nu=\frac{1}{3}$.

\begin{shaded}
There is an argument on whether a polymer molecule having a small $N$ undergoes the discontinuous transition. The argument might come from the theory of the partition function. The partition function can be related to the thermodynamic functions by the equation: $Z=\sum_{N}\exp(-A_{i}/kT)$ along with $A=G-PV$, while a finite sum of the exponential function cannot produce the singularity. Then they argue that the discontinuous coil-globule transition cannot occur for a small $N$. However, the prediction that a small system obeys the sudden conformational change is by no means inconsistent with the theory of the partition function. Note that there is no infinite system in reality; there is no mathematical singularity in our world. In this point of view, the gas-liquid transition of one mole of CO$_{2}$ which Andrews\cite{Andrews} once observed, equally to the coil-globule transition of a short molecule, is an event of a finite system.  To make the point clearer, it will be sufficient to cite an example of the Fourier series expansion: a finite sum of trigonometric functions can well approximate the discontinuous electric pulse. By analogy, it is possible that the finite sums of the partition functions well describe (seemingly abrupt) physical phenomena such as the gas-liquid phase transition, the coil-globule transition and so forth, and also can approximate with sufficient precision the true  singularity. The main point is that the discontinuity and the abruptness are our artificial construction in mathematics.
\end{shaded}

Then let us investigate concentrated regions.

\section{Branched Polymers in Concentrated Solution}
As has been shown previously\cite{Kazumi}, the density inhomogeneity inside a chain molecule rapidly diminishes with increasing number $N$ of segments, and the density flatness advances. As a result, with increasing $N$, a linear molecule must approach necessarily an unperturbed coil ($\alpha=1$) in all concentration ranges, except for the dilution limit ($\bar{\phi}\rightarrow 0$), namely, it behaves, for $N\rightarrow\infty$, as
%%%%%%%%%%%%%%%%%%% 24
\begin{equation}
\langle s_{N}^{2}\rangle^{1/2}\propto N^{\nu}\hspace{3mm}
\begin{cases}
\nu_{good}=\frac{3}{5} & \mbox{for}\hspace{3mm} \bar{\phi}\rightarrow 0\\[3mm]
\nu_{good}=\frac{1}{2} & \mbox{for}\hspace{3mm} 0<\bar{\phi}\le1
\end{cases}\label{AlD-24}
\end{equation}
which reveals that the exponent, $\nu$, has a critical nature; there is no intermediate value between $\nu_{\bar{\phi}\rightarrow 0}=\frac{3}{5}$ and $\nu_{melt}=\frac{1}{2}$. In the entire interval of $0<\bar{\phi}\le1$, $\nu$ takes the single exponent, $\frac{1}{2}$. In other words the exponent changes abruptly, through the infinitesimal change of $\bar{\phi}$, from $\nu=\frac{3}{5}$ in the dilution limit to $\frac{1}{2}$ in the finite concentration.

%%%%%%%%%%%%%%%%%% Fig-4
\begin{figure}[h]
\begin{center}
\begin{minipage}[t]{0.46\textwidth}
\begin{center}
\includegraphics[width=7.6cm]{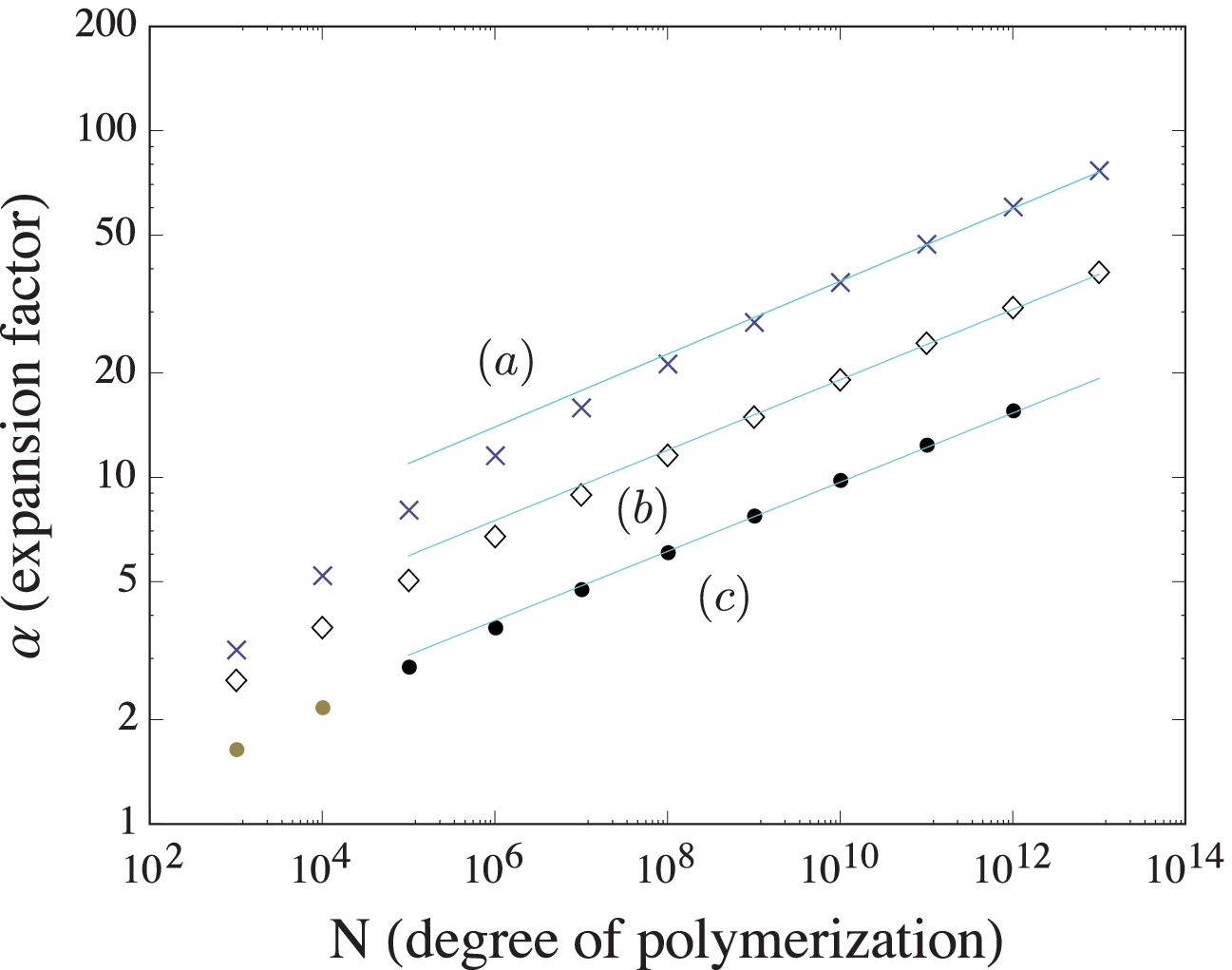}
\end{center}
\vspace{-2mm}
\caption{Expansion factor as against $N$ for the hypothetical branched polymer solution ($V_{1}=160\text{\AA}^{3}$, $V_{2}=387\text{\AA}^{3}$, $\bar{l}=10\text{\AA}$, $\chi=0$). The plot points were calculated according to Eq. (\ref{AlD-16}) for (a) $\bar{\phi}=0.01$, (b) $\bar{\phi}=0.1$, and (c) $\bar{\phi}=1\, (melt)$.}\label{AlphaGood}
\end{minipage}
\hspace{10mm}
%%%%%%%%%%%%%%%%%% Fig-5
\begin{minipage}[t]{0.46\textwidth}
\begin{center}
\includegraphics[width=7.8cm]{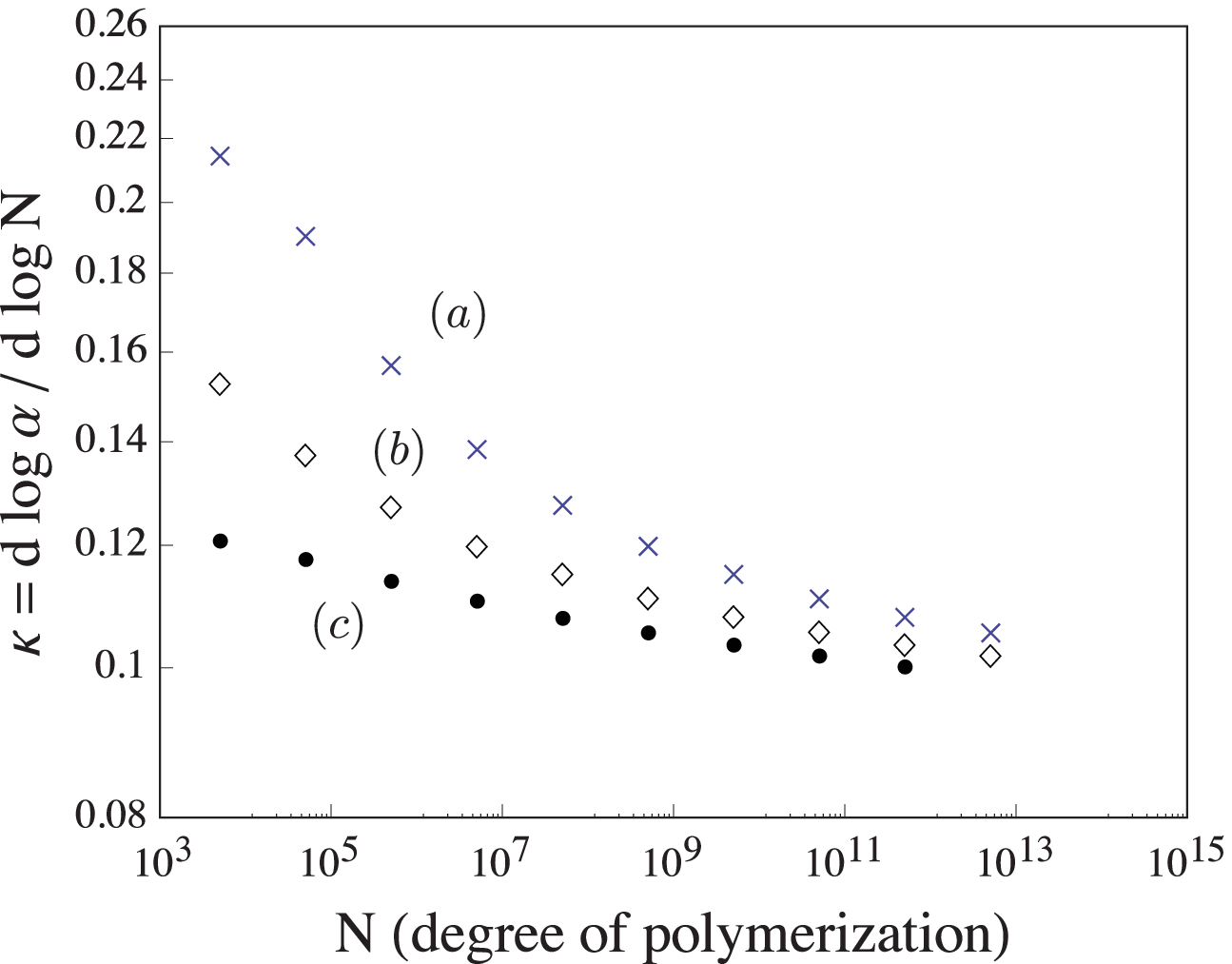}
\end{center}
\vspace{-2mm}
\caption{Variation of the gradient $\kappa=d\log\alpha/d\log N$ as a function of $N$ for the hypothetical branched polymer system ($V_{1}=160\text{\AA}^{3}$, $V_{2}=387\text{\AA}^{3}$, $\bar{l}=10\text{\AA}$, $\chi=0$): (a) $\bar{\phi}=0.01$, (b) $\bar{\phi}=0.1$, and (c) $\bar{\phi}=1\, (melt)$.}\label{KappaSlope}
\end{minipage}
\end{center}
\vspace*{-4mm}
\end{figure}

Then the question arises naturally as to whether branched polymers have the same critical nature. To answer this question we have examined the $\Bar{\phi}$ dependence of $\nu$ for a model branched polymer system ($V_{1}=160\text{\AA}^{3}$, $V_{2}=387\text{\AA}^{3}$, $\bar{l}=10\text{\AA}$, $\chi=0$, $d=3$). The numerical results are illustrated in Figs. \ref{AlphaGood} and \ref{KappaSlope} for $\bar{\phi}=0.01, 0.1$ and $1$\,\footnote{\,We have neglected $\Hat{\mathscr{J}}^{2}$ term in Eq. (\ref{AlD-16}), since $\Hat{\mathscr{J}}^{1}\gg\Hat{\mathscr{J}}^{2}$ in this simulation.}. It is seen that the scaling relation, $\alpha\propto N^{\kappa}$, is poorly convergent for all $\Bar{\phi}\,'s$. In the interval $N=[10^{3}, 10^{13}]$ simulated, the gradient, $\kappa=d\log\alpha/d\log N$, still doesn't reach the asymptotic limit. On the other hand, Fig. (\ref{KappaSlope}) suggests strongly that in the entire interval of $0<\bar{\phi}\le1$, the gradient converges on a common value that lies somewhere in the interval $\frac{1}{12}\le\kappa< 0.1$. Since $\nu=\nu_{0}+\kappa$, this leads to $0.33\cdots\le\nu <0.35$. It seems that ``0.35'' is close to $0.33\cdots$, the value evaluated in the preceding paper\cite{Kazumi}.

\section{Conclusion}
In summary, combining with the previous findings\cite{Kazumi}, we can compare various $\nu$ values for branched systems with the corresponding values for linear polymers (see Table \ref{AlD-Table2}). The results in Table \ref{AlD-Table2} suggest the exponent $\nu$ has a critical nature. It varies from phases to phases abruptly; for instance, it changes abruptly from $\nu_{good}=\tfrac{1}{2}$ in the dilution limit to $\nu_{melt}=\tfrac{1}{3}$ in the melt state; there is no intermediate value between them.

 %%%%%%%%%%%%%%%%%% Table 2
 \begin{table}[H]
 \centering
  \begin{threeparttable}
    \caption{Values of the exponent, $\nu$, for hypothetical branched polymer systems, $\langle s_{N}^{2}\rangle^{1/2}\propto N^{\nu}$ ($d=3$), and comparison with those for linear polymers}\label{AlD-Table2}
  \begin{tabular}{l l c c}
\hline\\[-1.5mm]
system & \hspace{10mm} exponent $\nu$ &  & \hspace{9mm} concentration \,\,\,\,\\[2mm]
\hline\\[-1.5mm]
branched polymer  & \hspace{10mm} $\nu_{globule}=\frac{1}{3}$  & \hspace{5mm}$T<\Theta$ & \hspace{5mm} dilution limit\\[1.5mm]
& \hspace{10mm} $\nu_{\Theta}=\frac{7}{16}$  & \hspace{5mm}$T=\Theta$ & \hspace{5mm} dilution limit\\[1.5mm]
& \hspace{10mm} $\nu_{good}=\frac{1}{2}$  & \hspace{5mm}$T>\Theta$ & \hspace{5mm} dilution limit\\[2.5mm]
& \hspace{10mm} $\nu_{good}=\frac{1}{3}$  & & \hspace{5mm} $0<\bar{\phi}\le1$\\[1.5mm]
\hline
\hline\\[-1.5mm]
linear polymer  & \hspace{10mm} $\nu_{globule}=\frac{1}{3}$  & \hspace{5mm}$T<\Theta$ & \hspace{5mm} dilution limit\\[1.5mm]
& \hspace{10mm} $\nu_{\Theta}=\frac{1}{2}$  & \hspace{5mm}$T=\Theta$ & \hspace{5mm} dilution limit\\[1.5mm]
& \hspace{10mm} $\nu_{good}=\frac{3}{5}$  & \hspace{5mm}$T>\Theta$ & \hspace{5mm} dilution limit\\[2.5mm]
& \hspace{10mm} $\nu_{good}=\frac{1}{2}$  & & \hspace{5mm} $0<\bar{\phi}\le1$\\[2mm]
\hline\\[-6mm]
   \end{tabular}
  \end{threeparttable}
  \vspace*{4mm}
\end{table}

%%%%%%%%%%%%%%%%%% References

\end{document}